# Reversible Single Spin Control of Individual Magnetic Molecule by Hydrogen Atom Adsorption

Liwei Liu[1†], Kai Yang[1†], Yuhang Jiang[1†], Boqun Song[1], Wende Xiao[1], Linfei Li[1], Haitao Zhou[1], Yeliang Wang[1], Shixuan Du[1], Min Ouyang[2], Werner A. Hofer[3,1], Antonio H. Castro Neto[4], and Hong-Jun Gao[1*]

[1] Institute of Physics, Chinese Academy of Sciences, P.O. Box 603, Beijing 100190, China.

[2] Department of Physics and Center for Nanophysics and Advanced Materials, University of Maryland, College Park, MD 20742, USA.

[3] Department of Physics, The University of Liverpool, Liverpool L69 3BX, UK.

[4] Graphene Research Centre, Department of Physics, National University of Singapore, 117542, Singapore.

†These authors contribute equally to this work.

*To whom correspondence should be addressed. Email: hjgao@iphy.ac.cn.

**The reversible control of a single spin of an atom or a molecule is of great interest in Kondo physics and a potential application in spin based electronics. Here we demonstrate that the Kondo resonance of manganese phthalocyanine molecules on an Au(111) substrate have been reversibly switched off and on *via* a robust route through attachment and detachment of single hydrogen atom to the magnetic core of the molecule. As further revealed by density functional theory calculations, even though the total number of electrons of the Mn ion remains almost the same in the process, gaining one single hydrogen atom leads to redistribution of charges within 3*d* orbitals with a reduction of the molecular spin state from S = 3/2 to S = 1 that directly contributes to the Kondo resonance disappearance. This process is reversed by a local voltage pulse or thermal annealing to desorb the hydrogen atom.**

Control over charge and spin states at the single molecule level is crucial not only for a fundamental understanding of charge and spin interactions but also represents a prerequisite for development of molecular electronics and spintronics[1-3]. While charge manipulation has been demonstrated by gas adsorption[4] and atomic manipulation[5], the reversible control of a single spin of an atom or a molecule has been challenging. Typically, atomic or molecular spin can be probed as a Kondo effect that manifests itself as a conductance anomaly at the Fermi level when it is coupled to a metallic system[6,7]. However, an effective method to manipulate molecular spin both individually and ensemble has been limited[8]. In this paper, we present the reversible control of a single spin in manganese phthalocyanine molecules on an Au(111) substrate. This process can be reversed by a local voltage pulse or thermal annealing to desorb the hydrogen atom, accompanied by a recovery of the molecular Kondo resonance.

Molecules with a single metal ion, including metal phthalocyanines (MPcs), metal porphyrins (MPs) and their derivatives, are ideal prototypes to study spin related phenomena due to their versatility and the tunability of charge and spin properties[9,10]. When magnetic MPcs or MPs are placed on metal surfaces, the Kondo resonance induced by the interaction between a localized spin of the molecule and conduction electrons from the substrate manifests itself as a narrow, pronounced peak in the density of states close to the Fermi level. While many attempts have been applied to govern the Kondo process, including conformation variation[11-15], alteration of adsorption site[16,17], quantum size effects[18], ligand attachment[19], molecular assembly[20] and atomic doping[21], most are either irreversible or cannot manipulate the Kondo effect at both single-molecule scale and ensemble; an effective method of molecular spin manipulation has been limited. We show that spin state of a single manganese phthalocyanine (MnPc) molecule on the Au(111) can be reversibly switched by chemical absorption and desorption of a single hydrogen atom. Our novel reversible spin control scheme can be easily realized at both ensemble of molecules and single-molecule level, which opens up new avenue of broader applications based on molecular spin state.

Figure 1a shows a typical low temperature scanning tunneling microscopy (LT-STM) image of MnPc molecules on an Au(111) substrate, in which every MnPc molecule possesses three inherent unpaired electrons with a total spin of $S = 3/2$. After deposition of 0.05 monolayer of MnPc onto a $22 \times \sqrt{3}$ herringbone reconstructed Au(111) surface, isolated MnPc molecules preferentially occupy the elbow and face centered cubic (fcc) sites. The STM image unambiguously confirms an intact molecular structure after thermal deposition with exhibition of a protruding "cross" feature with four-fold symmetry[16,18]. After the MnPc/Au(111) sample was dosed with 30 Langmuir (L) of $H_2$ at room temperature (RT) and subsequently cooled down to 4.2 K, all molecules retained the cross feature but some showed a depression at the center. After extensive $H_2$ dosage, all of the original MnPc molecules can be turned into molecules with a depression at the center (Fig. 1b and Supplementary Fig. 1).

The variation of the molecular morphology after the $H_2$ gas dosage can be attributed to the chemical adsorption of a single hydrogen atom onto the MnPc molecule, according to key control experiments summarized as follows: (1) This change of molecular morphology is reversible. We observed that by applying a positive voltage pulse (e.g. 2.0 V for one second with feedback loop off) the bright protrusion feature of the MnPc molecules can be recovered on a depression. This behavior is consistent with a tip induced detachment[4] of adsorbates from the MnPc molecule. Alternatively, the depression feature can be switched back to a bright protrusion by thermal annealing of samples at 500 K for 10 mins. (2) The apparent height of the molecule remains unchanged when the $H_2$ is dosed at 4.2 K instead of RT, even with very high dosage capacity (1000 L). We observed that intact $H_2$ molecules can only reside around the MnPc molecules without modification of the intrinsic molecular structure (Supplementary Fig. 2). This suggests that the interaction between the $H_2$ and the MnPc molecules is weak and absorption of molecular hydrogen cannot be the origin of the process demonstrated in the Fig. 1. (3) A threshold voltage of +1.28 V is required in order to detach the adsorbate from the host MnPc molecule (Supplementary Fig. 3). This threshold voltage suggests the binding between adsorbate and the molecule is chemical adsorption rather than physical adsorption, which demonstrates that the absorption of the $H_2$ cannot be the origin of the process shown in the Fig. 1. (4) Lastly we also carried out one control experiment by direct dosage of atomic hydrogen gas[22] (Supplementary Section 2.), and similar features of the molecular morphology change were also obtained. From these key experiments we conclude that the depression feature after dosage of hydrogen gas is due to adsorption of atomic hydrogen (labeled as "H-MnPc").

We explored and compared in detail the difference of electronic structure between the MnPc and the H-MnPc states by measuring differential conductance (*dI/dV*) spectra. As shown in the upper red curve in Fig. 1c, the *dI/dV* spectra acquired at the center of the original MnPc molecules consistently show a pronounced step shaped feature at zero bias, regardless of the molecular adsorption configurations and locations on the Au terraces. Such step shaped feature in *dI/dV* spectra is very sharp and close to the Fermi level. This spectra feature manifests linear splitting at the presence of magnetic field (Fig. 1d) as well as temperature dependence (Supplementary Fig.8), and thus it can be safely attributed to the Kondo resonance[6,7,23,24]. In contrast, the *dI/dV* spectra of the H-MnPc state (middle blue curve in Fig. 1c) are featureless in this energy range. Importantly, once the H-MnPc state was switched back to the MnPc state, its electronic structure including the Kondo resonance can be fully recovered (bottom red curve in Fig. 1c) in addition to the recovery of the topographic feature.

Further insight into the reversible switching of the Kondo resonance between the MnPc and the H-MnPc states can be gained by density functional theory (DFT) calculations (see Method for more details). Topographic features of both the MnPc and H-MnPc states can be well reproduced by STM simulations by integrating electron density from -0.4 eV to the Fermi level according to the Tersoff-Hamann approximation[25]. The projected density of states (PDOS) shows that the bright protrusion at the molecular center of MnPc is due to the *d*-orbital character of the Mn ion near the Fermi level (Fig. 2a, and Supplementary Fig. 5). The calculated binding energy of 1.75 eV between the hydrogen atom and MnPc/Au(111) indicates that H on MnPc/Au(111) is of a chemical adsorption with binding energy in the same range of our experimental value. Therefore, both experimental results and theoretical calculations suggest that attaching and detaching a single hydrogen atom from the central Mn ion is the physical origin of spin switching process.

An analysis of the electronic structure of MnPc and H-MnPc reveals that the number of *d*-orbital electrons of the Mn ion remains almost unchanged. Thus the additional electron contributed by the hydrogen atom is donated to the Pc skeleton. However, the effective charges in individual *d* orbitals are redistributed. Before hydrogen adsorption, the Mn ion in the MnPc molecule has a net spin of S=3/2. This is due to two partially paired orbitals, $d_{xz}$ and $d_{yz}$, and two unpaired orbitals, $d_{xy}$ and $d_{z^2}$ (Fig. 2b). The Kondo resonance observed in the *dI/dV* spectra in this case originates from an exchange interaction between the localized spin of the magnetic

Mn ion and the conducting electrons of the Au(111) substrate[26]. After atomic hydrogen decoration (i.e., in the H-MnPc state), the spin polarization decreases in the $d_{z^2}$ orbital: Before adsorption the spin-up charge in this state is close to one while the spin-down charge is close to zero. After adsorption the charge in both spin states is close to 1/2. This leads to a decrease of the net spin of the Mn ion from S=3/2 to S=1 (Fig. 2c). Because the Kondo effect strongly depends on the coupling between the localized spin and the conducting electrons, a reduced molecular spin state directly contributes to the suppression of the Kondo effect (Supplementary Section 4).[27,28] A schematics of manipulation of the molecular Kondo effect by hydrogen adsorption and desorption is illustrated in Fig. 2d.

In principle, the appearance or disappearance of the Kondo resonance on this magnetic molecule can be seen as a single bit of information. This may have implications for information recording and storage at the ultimate molecular limit[10-12,29-31], when combined with STM capability of atomic manipulation of molecular adsorption and desorption. We explore this avenue by investigating an ordered molecular array. Fig. 3a shows topography and *dI/dV* mapping acquired simultaneously from a 3 × 4 molecular array. All the MnPc molecules within the array are initialized by converting them to the H-MnPc state through the adsorption of hydrogen atoms, with evidence of featureless *dI/dV* mapping (i.e. absence of the Kondo resonance), while the four-lobe topographic structure of individual molecules remains (Fig. 3a). By applying voltage pulses on selected H-MnPc molecules, we can precisely address an individual molecule within the array and the H-MnPc molecules can be converted back to the MnPc state one-by-one. As a result, a pattern of confined dark dot feature corresponding to the Kondo resonance of the MnPc molecules can be created in the *dI/dV* mapping (Fig. 3b). Importantly, this process can be extended to an ultimate close-packed array with the highest molecular density (Supplementary Fig. 6 and Supplementary Fig. 7), with one example shown in the Fig. 3c. While the inter-molecular spacing is as small as 1.4 nm in a close-packed array, all the dark-dot features in the *dI/dV* mapping remain localized at the center of the MnPc molecules, and the *dI/dV* spectra of the Kondo switching recorded in the molecular assembly show no difference from our observation of isolated molecules (Supplementary Fig. 7), regarding characteristic parameters such as the Fano factor (*q*) and the Kondo temperature ($T_K$). Correspondingly, no change occurs in the molecular framework during the spin manipulations within the molecular assembly, in contrast to previously investigated systems, which involve conformational changes at the periphery of the molecule[11,13,14]. A storage system based on the molecular spin states thus describes the ultimate limit of information storage. Furthermore, all the MnPc molecules in the arrays exhibit consistent Kondo features, suggesting a site-independent Kondo resonance and the inter-particle interactions can be weak enough to become negligible in the process. (Fig. 1 and Supplementary Fig. 7). Even more importantly, such molecular spin switching can be consistently achieved back-and-forth for many times with no observable change of the Kondo features. To substantiate the robustness of the spin switching process, we have quantified both $T_K$ and *q* of MnPc state from a Fano fitting (Supplementary Fig. 8) and monitored both parameters during the spin switching process by repeatedly absorption and desorption of H-atom, with results summarized in the Fig. 3d. That both, the $T_K$ and *q* values in the spin manipulation cycles remain constant supports the robustness of this process. Atomic localization feature of the Kondo resonance switching makes it possible to create more complex patterns based on molecular spin state, including one example of a digitalization from (000) to (111) that can be fabricated from a larger 3 × 8 molecular array (Supplementary Fig. 9).

The ability to reversibly tune the Kondo effect at the molecule level by gaining or losing a single hydrogen atom creates a new class of simplest molecular Kondo system in which the spin screen can be manipulated directly through the control of the local atomic environment. Our demonstration of reversible molecular spin control by chemical stimuli is different from a few recent works in that it offers an in-situ single molecule characterization and control instead of ensemble measurement[32] and selection of noble metallic Au substrate can further avoid complex spin texture of substrate due to Rashba type spin-split pair surface states of a semimetal of such as Bi[33,34]. Even more importantly, our scheme can allow not only ensemble spin manipulation (such as Kondo ON/OFF switching of all molecules by thermal annealing and $H_2$ gas dosage, respectively) but also control at the single-molecule level by atomic manipulation of single H atom that has not been achieved so far[21]. It thus opens up a variety of avenues. Our approach can be readily extended to more complex magnetic molecules. The atomically precise engineering of molecular spin states provides a unique test ground for probing magnetic interactions at the single-molecule level. For example, the magnetism in transition-metal ions based magnetic molecules arises from unpaired spins residing in the *d*-orbitals. Controlled coupling of its net spins with another unpaired spin of a well-defined hydrogen atom provides the simplest conceivable systems potentially exhibiting quantum critical behavior. Our demonstration of robust Kondo switching in a molecular array could also be applied as nonvolatile classical information recording and storage at the ultimate molecule limit without the requirement of molecular conformation change. Future design of more complex molecular framework might offer a way to enable and tailor coherent spin coupling through such as inter-molecular bridge[35].

# Figure Captions

**Figure 1 | Topography and *dI/dV* curves of the MnPc molecules on the Au(111) before and after hydrogen atom adsorption. a,** Topography of the MnPc molecules on the Au(111) before hydrogen addition, showing molecules with bright center. Scanning bias: -0.2 V; Tunneling current: 10 pA; Scale bar, 5 nm. (Inset) High-resolution image of a single molecule. Scale bar, 0.7 nm. **b,** Topography of the MnPc molecules after hydrogen atom decoration, showing some molecules with suppressed center. Scanning bias: -0.2 V; Tunneling current: 10 pA; Scale bar, 5 nm. (Inset) High-resolution image of a single molecule with suppressed center. Scale bar, 0.7 nm. **c,** Sequential variation of the *dI/dV* spectra recorded at the center of a MnPc molecule induced by the absorption and desorption of a single hydrogen atom. For comparison purpose spectra were vertically shifted 0.1 nA/V and 0.33 nA/V for middle and upper curves, respectively. **d,** *dI/dV* spectra acquired at the center of the MnPc molecule under external magnetic field at 0.4 K, showing the splitting of Kondo resonance. For clarity the successive spectra were vertically shifted by 0.1 nA/V.

**Figure 2 | STM simulation and PDOS of the MnPc/Au(111) before and after hydrogen adsorption a,** STM simulation of the MnPc/Au(111) and the H-MnPc/Au(111), confirming the topographic feature in Fig. 1**a** and **b**. **b,** and **c,** PDOS of $d_{xz}$, $d_{yz}$ and $d_{z^2}$ orbitals of Mn ion in the MnPc /Au(111) and the H-MnPc /Au(111), respectively, revealing that spin at the Mn ion is reduced from 3/2 to 1. **d,** Schematics of reversible control of molecular Kondo effect with adsorption (Kondo OFF) and desorption of hydrogen atom (Kondo ON).

**Figure 3 | Reversible spin switching in molecular arrays. a,** The *dI/dV* mapping (upper panel) and simultaneously acquired topography (lower panel) of a H-MnPc molecular array (Kondo OFF state). **b,** The *dI/dV* mapping (upper panel) and simultaneously acquired topography (lower panel) of the same molecular array as **a** but with a few selected molecules converted to the MnPc state (Kondo ON). For both **a** and **b,** the *dI/dV* mappings were taken at 6 mV. Note that the MnPc state (Kondo ON) can be erased and converted back to H-MnPc state (Kondo OFF) by adsorption of additional hydrogen. **c,** The *dI/dV* mapping (upper panel) and topography (lower panel) of a close-packed molecular array with pre-designed Kondo pattern by switching selected molecules from the H-MnPc to the MnPc states within the array. The *dI/dV* mapping was taken at 6 mV. **d,** The Fano factor *q* and Kondo temperature $T_K$ of Kondo ON states after multiple-cycles of spin switching, indicating the robustness of the spin manipulation process.

## METHODS SUMMARY

### Experimental method

Experiments were carried out in an ultrahigh vacuum (base pressure of $1\times10^{-10}$ mbar) LT-STM system (Unisoku), equipped with standard surface processing facilities. An atomically flat Au(111) surface was prepared by repeated cycles of sputtering with argon ions and annealing at 800 K. Commercial MnPc molecules (Sigma-Aldrich, 97% purity) were sublimated from a Knudsen-type evaporator after thermal purification, while the Au(111) substrate was held at room temperature. One monolayer refers to a close-packed MnPc layer covering the whole Au(111) surface, as examined by STM (Supplementary Fig. 6). STM images were acquired in the constant-current mode with bias voltage applied to the sample. Spectroscopic measurements were performed by using a lock-in technique with a 0.5 mV$_{rms}$ sinusoidal modulation at a frequency of 973 Hz. Most images and spectra were obtained with electrochemically etched tungsten tips at 4.2 K, but the magnetic field dependent measurement was performed at 0.4 K. Measurements of local spectroscopy were electronically calibrated by performing *dI/dV* measurements on clean Au(111) before and after measurements on molecules, based on the well-known Au(111) surface state [36]. The atomic H is generated by cracking H$_2$ molecules with hot tungsten filament, which is placed very close (~5 cm) to the sample.

### Calculation details

The DFT calculations were performed using Vienna ab-initio simulation package (VASP)[37]. Perdew-Wang 91 exchange correlation functionals and the projector augmented-wave method were employed[38]. The energy cutoff for plane-wave basis set was 400 eV. A *c*(7×8) supercell containing three layers of gold atoms was employed to model an isolated molecule on Au(111) substrate. To ensure that interactions between the periodic slabs through vacuum can be negligible, the slabs were separated by a vacuum gap of 20 Å. For geometry optimization, the bottom layer was fixed, while the adsorbates and the other metal layers were allowed to relax until the forces applied to the relaxed atoms were less than 0.01 eV/Å. A single Γ point was used in sampling the Brillouin zone due to the numerical limitations (See Supplementary Section 8).

## Acknowledgements

The authors would like to thank Wilson Ho for helpful discussion. G. Li, L. Meng, L.H. Yan, and X.M. Fei for experimental assistance. Work at IOP was supported by grants from National Science Foundation of China (Grant NSFC 20973196), National "973" projects of China, the Chinese Academy of Sciences, and SSC (Grant 973 2009CB929103). Work in Liverpool is supported by the EPSRC Car-Parinello consortium, grant No EP/F037783/1. M.O. acknowledged support from the ONR (award #: N000141110080). AHCN acknowledges DOE grant DE-FG02-08ER46512, ONR grant MURI N00014-09-1-1063, and the NRF award R-144-000-295-281.

## Author contributions

L.W.L., K.Y., and Y.H.J. contribute equally to this work. L.W.L., K.Y., Y.H.J., W.D.X., L.F.L., H.T.Z. and Y.L.W. performed STM measurements with the help of H.-J.G. B.Q.S., S.X.D. and W.A.H. conducted and supervised DFT calculations. L.W.L., S.X.D., M.O., W.D.X., W.A.H., A.H.C.N. and H.-J.G. analyzed the data and wrote the manuscript. H.-J.G. designed and coordinated the project. All authors discussed the results and commented on the manuscript.

**Figures**

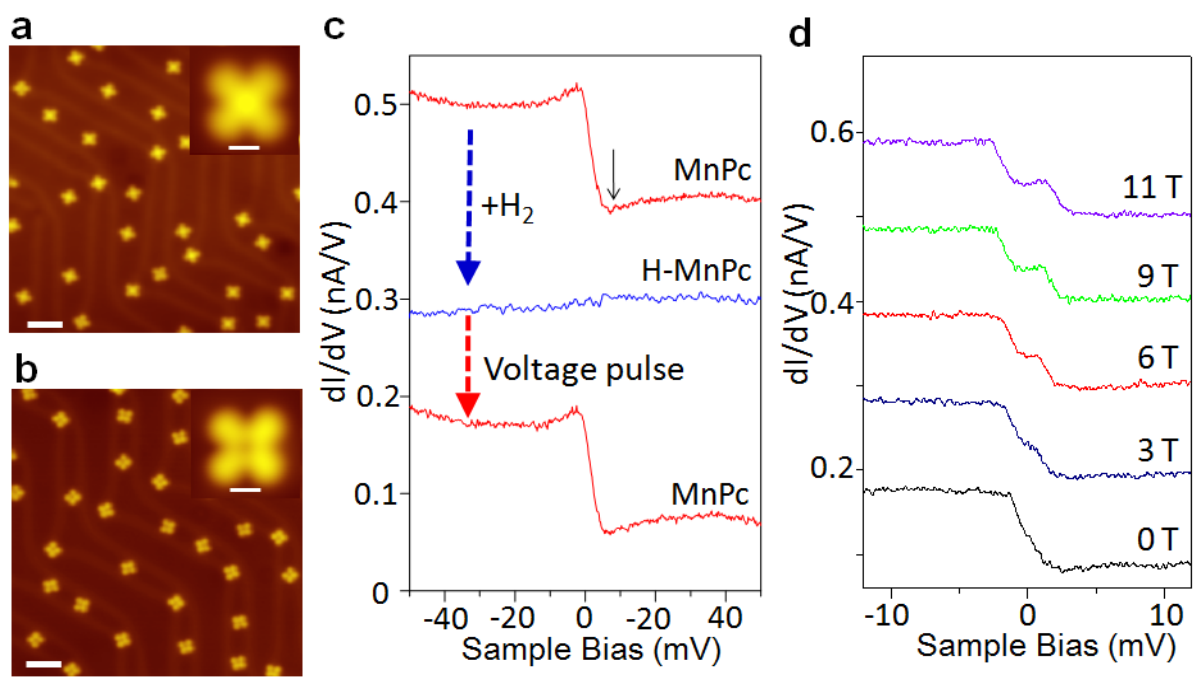

Figure 1

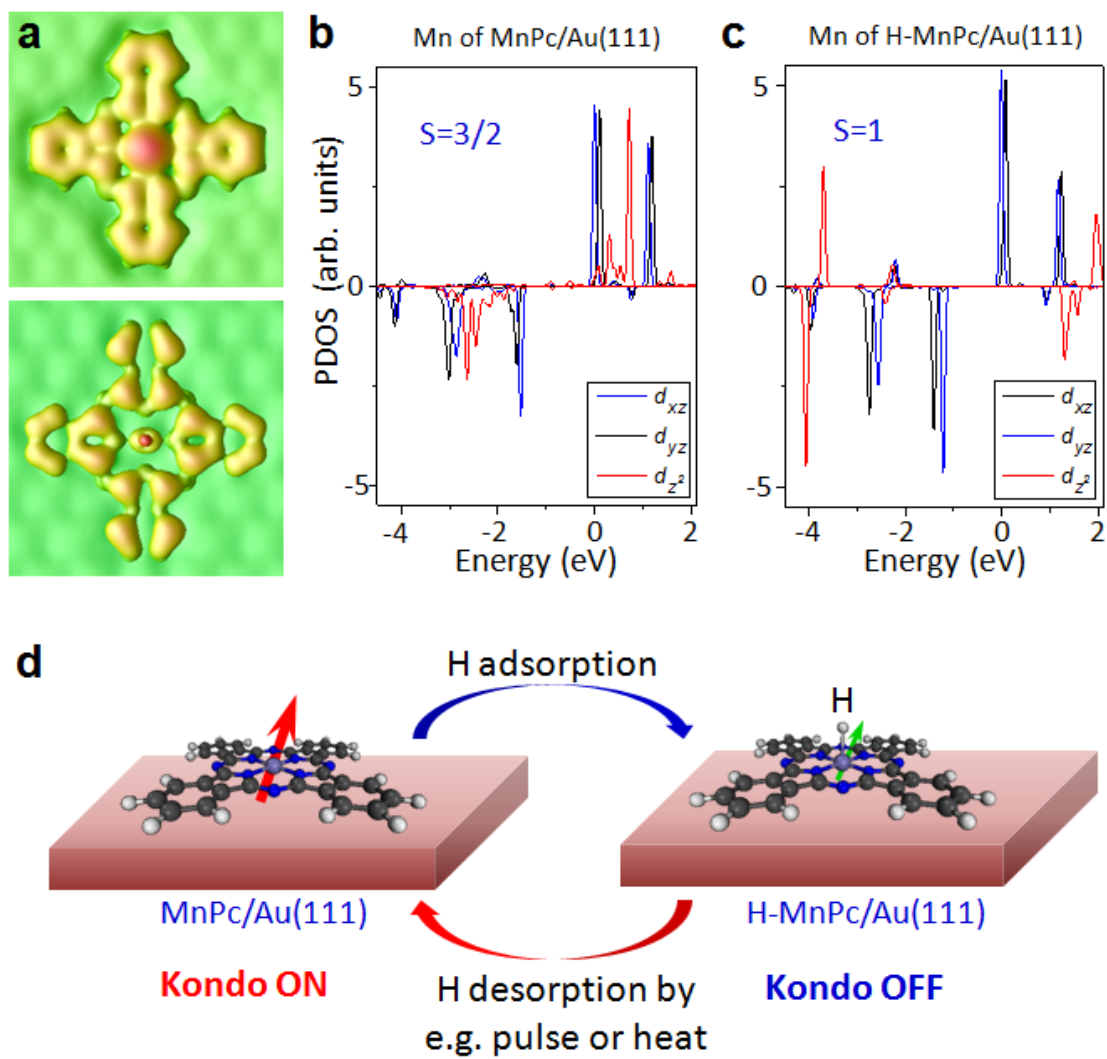

**Figure 2**

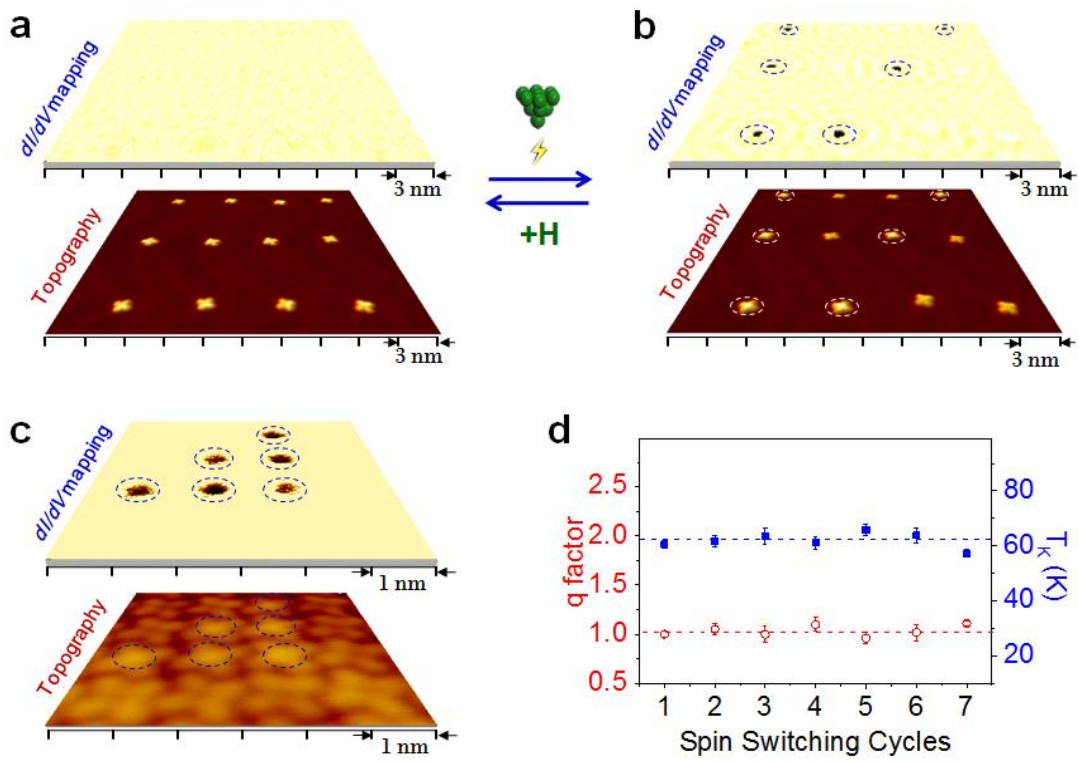

**Figure 3**

# Supplementary Information

# Reversible Single Spin Control of Individual Magnetic Molecule by Hydrogen Atom Adsorption

Liwei Liu[1†], Kai Yang[1†], Yuhang Jiang[1†], Boqun Song[1], Wende Xiao[1], Linfei Li[1], Haitao Zhou[1], Yeliang Wang[1], Shixuan Du[1], Min Ouyang[2], Werner A. Hofer[3,1], Antonio H. Castro Neto[4], and Hong-Jun Gao[1*]

[1] Institute of Physics, Chinese Academy of Sciences, P.O. Box 603, Beijing 100190, China.

[2] Department of Physics and Center for Nanophysics and Advanced Materials, University of Maryland, College Park, MD 20742, USA.

[3] Department of Physics, the University of Liverpool, Liverpool L69 3BX, UK.

[4] Graphene Research Centre, Department of Physics, National University of Singapore, 117542, Singapore.

**Contents**

†These authors contribute equally to this work.

*To whom correspondence should be addressed. Email: hjgao@iphy.ac.cn.

# 1. Conversion of the MnPc state to the H-MnPc state by $H_2$ dosage at room temperature

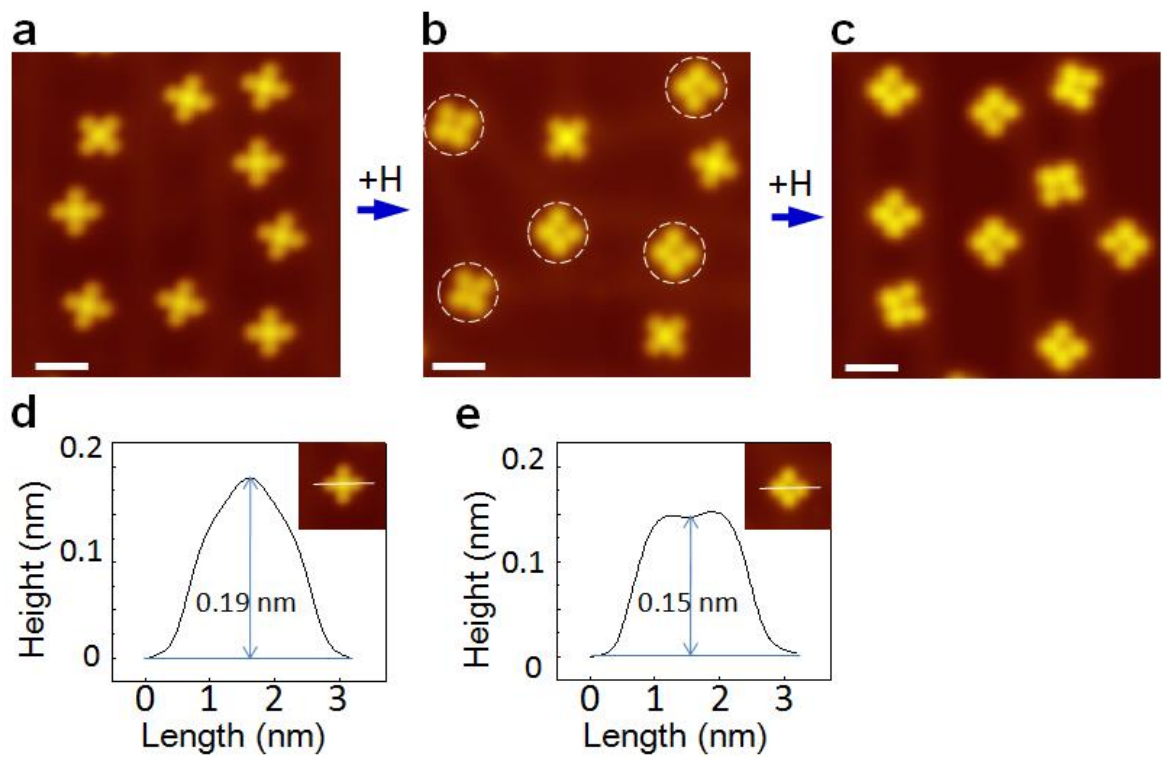

**Supplementary Figure 1** | Topography of the MnPc/Au(111) under different hydrogen dosage condition. **a,** The MnPc molecules on Au(111) surface before hydrogen dosage, showing a bright protrusion at the molecular center for all molecules. **b,** The MnPc molecules after low dosage of $H_2$ gas at room temperature, showing that some molecules (the molecules highlighted by dashed circles) are converted to the H-MnPc state with evidence of slightly dark center. **c,** All of the MnPc molecules are converted to the H-MnPc state under high dosage of hydrogen gas at room temperature. Scanning bias: -0.2 V; Tunneling current: 8 pA; Scale bar, 2 nm. **d,** A typical line profile across center of the MnPc in the inset. Scanning bias: -0.2 V; Tunneling current: 8 pA; Image size: 4.2 nm × 4.2 nm. **e,** A typical line profile across center of the H-MnPc in the inset. Scanning bias: -0.2 V; Tunneling current: 8 pA; Image size: 4.2 nm × 4.2 nm.

## 2. $H_2$ adsorption and decomposition on the Au(111)

We systematically investigated effects of dosage of $H_2$ gas to the MnPc/Au(111) system by varying substrate temperature from 300 K to 4 K. The adsorption of the $H_2$ molecules onto the substrate was evident in the whole temperature range in the *dI/dV* spectra on a bare Au surface, which can be attributed to the $H_2$ molecular vibration (Supplementary Fig. 2g)[1]. However, there existed a clear threshold of substrate temperature for the formation of the H-MnPc state: sample temperature needed to be maintained above 120 K during deposition of the $H_2$ gas in order to convert the MnPc state to the H-MnPc state. For sample temperature below 120 K, no conversion from the MnPc to the H-MnPc was observed. Supplementary Fig. 2 shows one example of $H_2$ dosage at 4.2 K. All MnPc molecules were not converted to H-MnPc state as evidenced by the line-profile across the molecular center, even though they were accumulated by $H_2$ molecules with a nearest-neighbor spacing of ~0.38 nm which is consistent with bulk solid hydrogen (Supplementary Fig. 2f)[1]. Existence of this threshold temperature further supports our assignment of the H-MnPc state. In order to achieve the H-MnPc state by dosage of hydrogen gas, hydrogen molecules need to be decomposed into hydrogen atoms when they arrive at the Au surface. It has been demonstrated that hydrogen molecules can be decomposed to hydrogen atoms with assistance of catalysts (typically noble metals, including Au applied in this study) and under high enough substrate temperature[2,3] Such morphology change can also be induced by exposing MnPc/Au(111) sample directly to atomic H gas (by decomposing $H_2$ molecules using hot tungsten filament before deposition onto sample[3]) when the sample is held either at RT or at low temperature of 80 K, further supporting our assignment of the adsorbate to H atom.

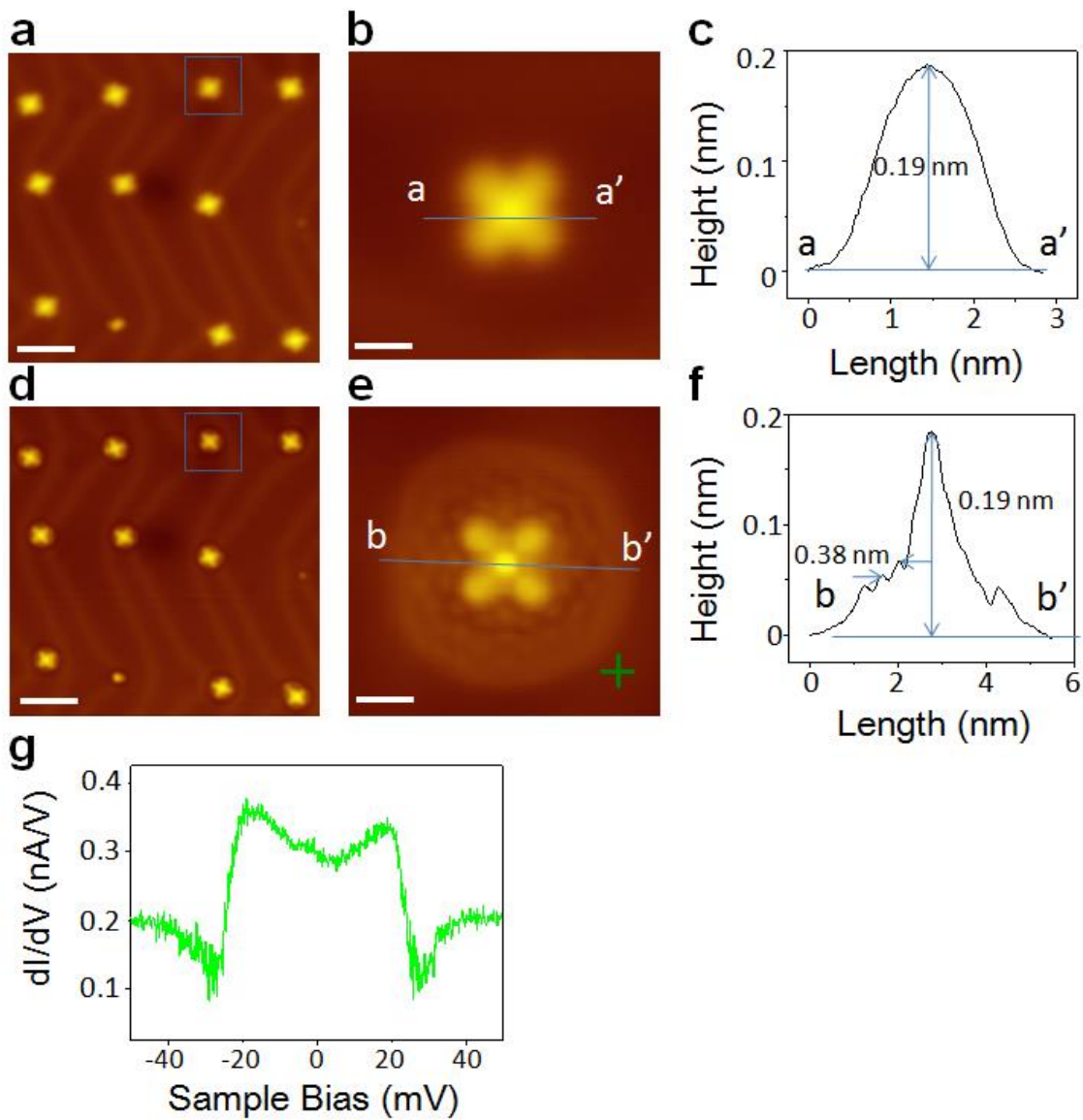

**Supplementary Figure 2 |** Topography of the MnPc/Au(111) before and after $H_2$ dosage at sample temperature of 4.2 K. **a,** Topography of MnPc/Au(111) before the $H_2$ dosage. **b,** Zoom-in topography of the MnPc molecule. Scanning bias: -0.2 V; Tunneling current is 8 pA for both **a** and **b**; Scale bar, 5 nm for **a** and 1 nm for **b**. **c,** Line profile across center of the MnPc in the **b**. **d** Topography of the MnPc/Au(111) after dosage of $H_2$ gas of 160 L, showing that the $H_2$ molecules accumulate around the MnPc and on bare Au substrate. **e,** Zoom-in topography of the MnPc molecule highlighted in **d** by a blue square. Scanning bias: -0.2 V for **d** and -0.1 V for **e**; Tunneling current is 8 pA for both **d** and **e**; Scale bar, 5 nm for **d** and 1 nm for **e**. **f,**

Line profile across center of the MnPc in **e**. **g,** The dI/dV curve of $H_2$ on the Au(111), acquired at the cross mark in **e**.

## 3. STM electric field effect on the H atom detachment process

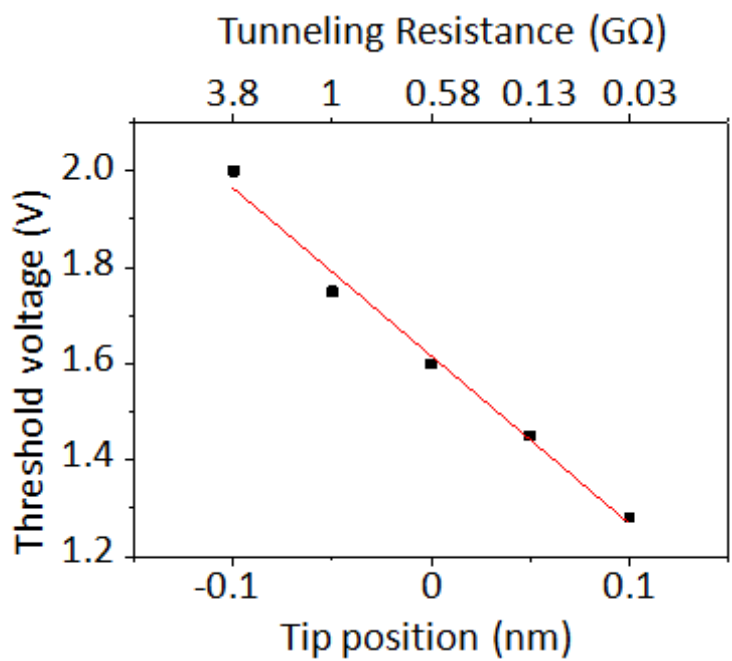

**Supplementary Figure 3 | The threshold voltage of H detachment from H-MnPc/Au(111).** The threshold voltage has linear dependence with relative position of the tip, suggesting that the H detachment is an electric field induced effect. A larger relative position means a decreased distance between the tip and molecules.

## 4. PDOS and STM simulation of the MnPc/Au(111) before and after hydrogen adsorption

We calculated and compared configurations of the MnPc and H-MnPc states by DFT calculations. The configurations as shown in Supplementary Fig. 4a are energetically stable after optimization. The adsorption energy of the MnPc on the Au(111) is found to be reduced by 163 meV after the H decoration, which suggests a reduced molecule-substrate interaction. In addition, our calculation shows that the separation between the molecular plane of the MnPc and the Au(111) surface increases from 3.1 Å to 4.0 Å after the H decoration (Supplementary Fig. 4b). Because Kondo scattering strongly depends on the coupling between the localized spin and the conducting electrons, both a decrease of molecular spin state (from S=3/2 to S=1, as discussed in the main text) and an increase of molecule-substrate spacing should contribute to the disappearance of the Kondo effect by the H-adsorption, but the reduction of net spin should play a dominant role[3]. Van der Waals interactions in this case will slightly decrease the distance between molecule and surface. However, they will not change the electronic structure or the overlap between molecular states and states of the substrate.

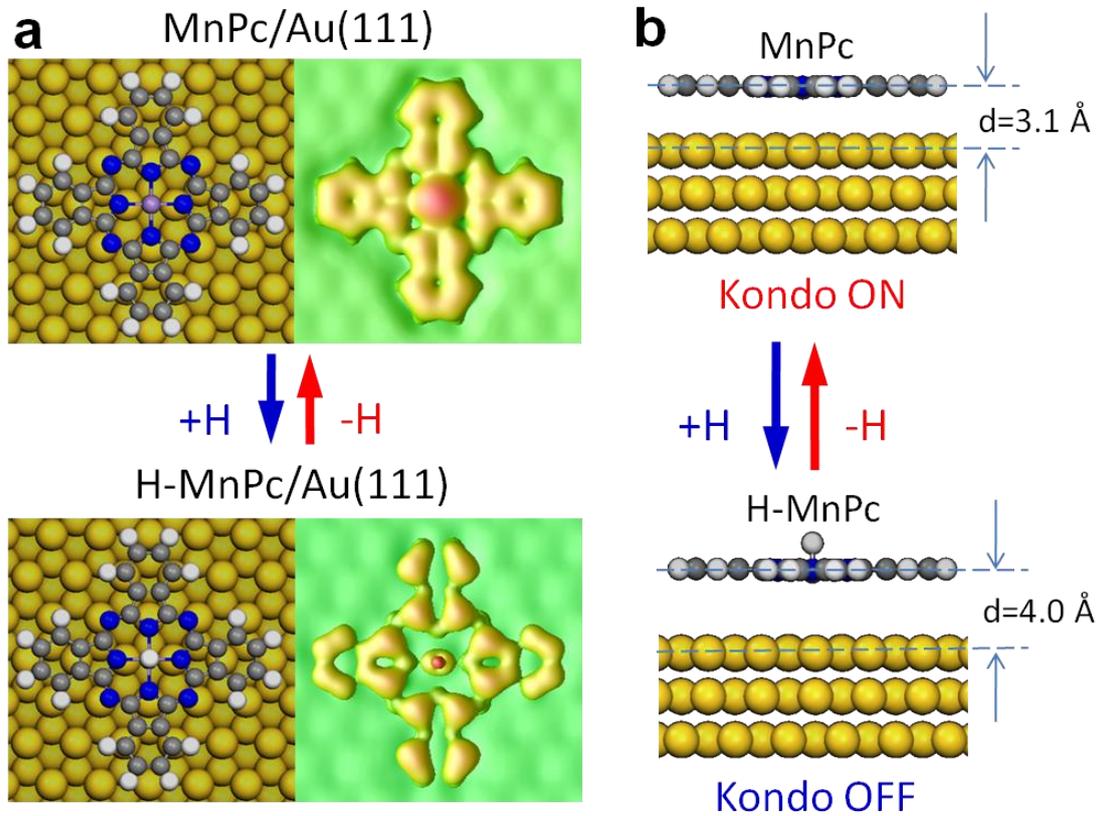

**Supplementary Figure 4 | a** and **b** PDOS of all d orbitals of Mn ion of the MnPc/Au(111) before and after H-adsorption, respectively.

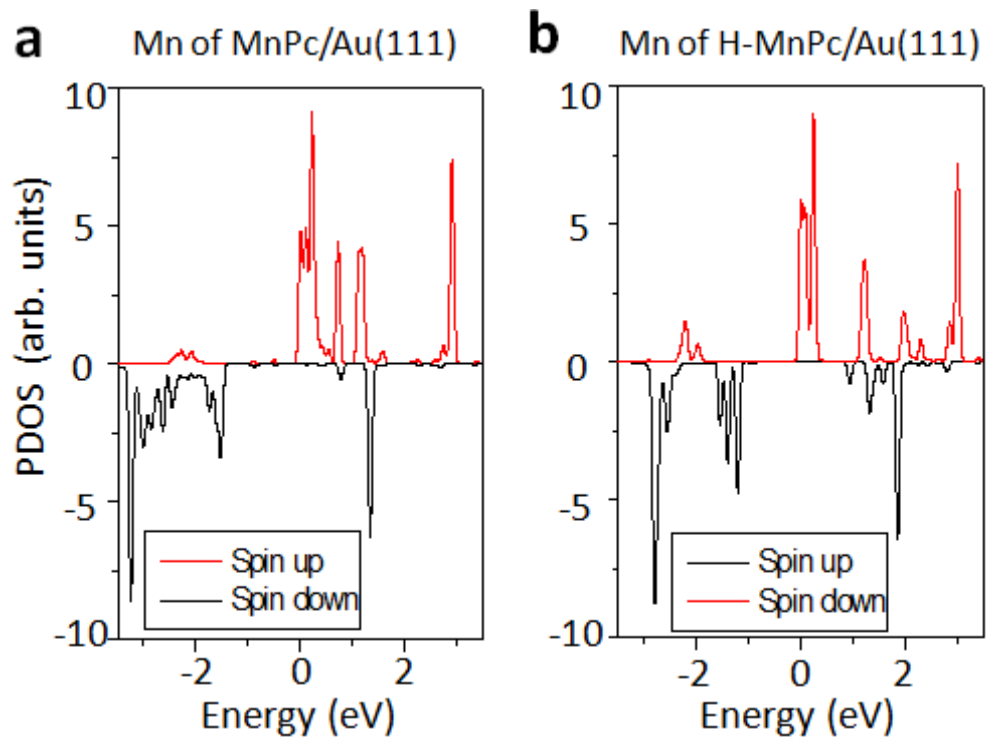

**Supplementary Figure 5 |** DFT calculation of molecular configurations of the MnPc and the H-MnPc molecules on Au(111). **a,** Top view of the MnPc (upper panel) and the H-MnPc molecules (lower panel) on the Au(111) surface. STM image simulations of corresponding adsorption configuration are provided at the right side of each ball-and-stick model. **b,** Side view of the MnPc (upper panel) and the H-MnPc (lower panel) molecules on the Au(111) surface.

We would like to discuss further about spin S=1 state after a single H-atom adsorption, which is also defined as Kondo OFF state in Figure 2. As compared with S=3/2 state, our current experiment cannot preclude a possibility that the S=1 state may lead to an underscreened Kondo effect[4] but with characteristic Kondo temperature below 0.4 K (i.e. current detection limit of our apparatus). In the future, mK-STM might be able to provide more insight of S=1 state as observed here. Since our spin manipulation experiments (Fig.3) were all performed at 4.2 K, it is safe to define Kondo On and Off states based on conductance features.

## 5. Spin switching in a close-packed molecular array

A close-packed molecular monolayer covering the whole Au(111) surface was measured by STM. As compared with isolated molecules it usually took longer time in order to convert all molecules within the array from the MnPc to the H-MnPc states under similar $H_2$ dosage condition. This observation is again consistent with our assignment of H-atom adsorbate: Because $H_2$ molecule needs to decompose and diffuse in order to form H-MnPc state this process could require more time in the situation of molecular array due to the restricted Au area.

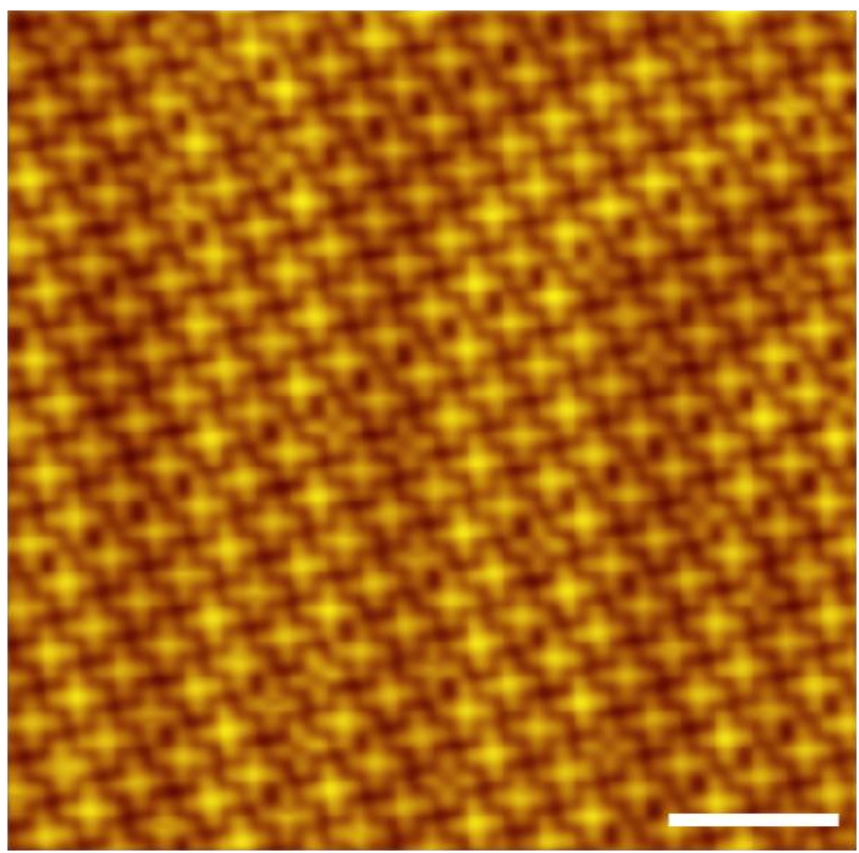

**Supplementary Figure 6 |** Large-scale STM image of close-packed molecular array with inter-molecular distance of 1.4 nm, combining a few molecules in the H-MnPc state. Scanning bias: -0.2 V; Tunneling current: 10 pA; Scale bar, 4 nm.

In order to evaluate effects of possible exchange interaction between neighboring molecules on the Kondo switching process, we investigated a close-packed monolayer sample. The Supplementary Fig. 6 and 7a show one example with some of the MnPc molecules that were already converted to the H-MnPc state (e.g., molecule highlighted by a blue dashed circle). This H-MnPc molecule was subsequently converted back to the MnPc state by applying a voltage pulse (red dashed circle in the Supplementary Fig. 7b). For comparison, the *dI/dV* spectra corresponding to different molecular states highlighted by blue, red and black dashed circles were presented and compared in the Supplementary Fig. 7c, respectively. We have observed no significant difference of Kondo ON and OFF states between close-packed molecular array and isolated molecules in Fig. 1c. This suggests that the effect of exchange interaction among neighboring molecules can be neglected[5,6]. Importantly, the fact that reproducible spin switching process of molecules within a close-packed array can be achieved and shows no difference from that of isolated molecules provides a proof-of-principle for future applications in ultrahigh density information storage and operation at the ultimate molecular limit.

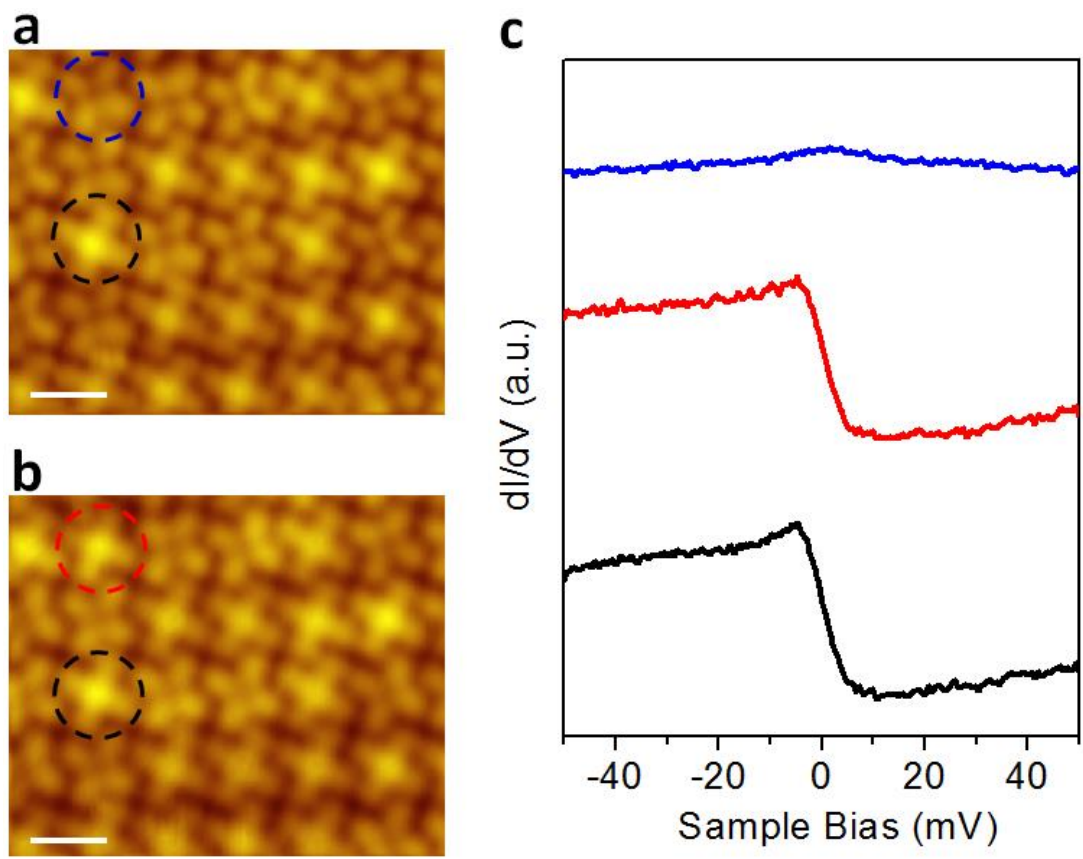

**Supplementary Figure 7 |** Topography and *dI/dV* curves of a 6 × 5 close-packed molecular array on Au(111) before and after hydrogen desorption. **a,** Topography of a

close-packed molecular array with a few molecules in the H-MnPc state. Scanning bias: -0.2 V; Tunneling current: 10 pA; Scale bar, 2 nm. **b,** Topography taken after a voltage pulse (2 V, 1 s) applied to a selected H-MnPc molecule (highlighted by a blue dashed circle in **a**. The selected H-MnPc molecule could be converted to the MnPc after the pulse (red dashed circle) without affecting neighboring molecules. Scanning bias: -0.2 V; Tunneling current: 10 pA; Scale bar, 2 nm. **c,** The *dI/dV* spectra at the center of a molecule before (blue curve) and after (red curve) hydrogen desorption by applying a voltage pulse. For comparison, a *dI/dV* spectrum (black curve) measured at the center of a nearby MnPc molecule (black dashed circle) is also provided. There is no difference of Kondo feature between red and black curves.

# 6. Kondo resonance with temperature and a typical fitting procedure for determining $q$ and $T_K$ of Kondo resonance

To analyze the Kondo resonance in the switching process quantitatively, we fit the *dI/dV* line shape with a Fano function[7],

$$\frac{dI}{dV}(V) = A \cdot \frac{(\varepsilon' + q)^2}{1 + \varepsilon'^2} + B \qquad (1)$$

where $\varepsilon' = (eV - \varepsilon_0)/\Gamma$, $A$ is the amplitude coefficient, $B$ is the background, $\varepsilon_0$ is the energy shift of the resonance from the Fermi level, $q$ is the Fano factor, $\Gamma \sim k_B T_K$ is the half width of the resonance, and $T_K$ is the Kondo temperature. By fitting to the Fano function, we obtain an average value of $q = -1.03 \pm 0.05$ and the $T_K \sim 61.9 \pm 2.7$ K for the Kondo ON state (Supplementary Figure 8a).

The temperature dependence of Kondo resonance was also measured on the monolayer MnPc on Au(111). As shown in Supplementary Figure 8b, the Kondo resonance disappears at 77K, which is consistent with our estimation of $T_K$ at ~ 61.9 K.

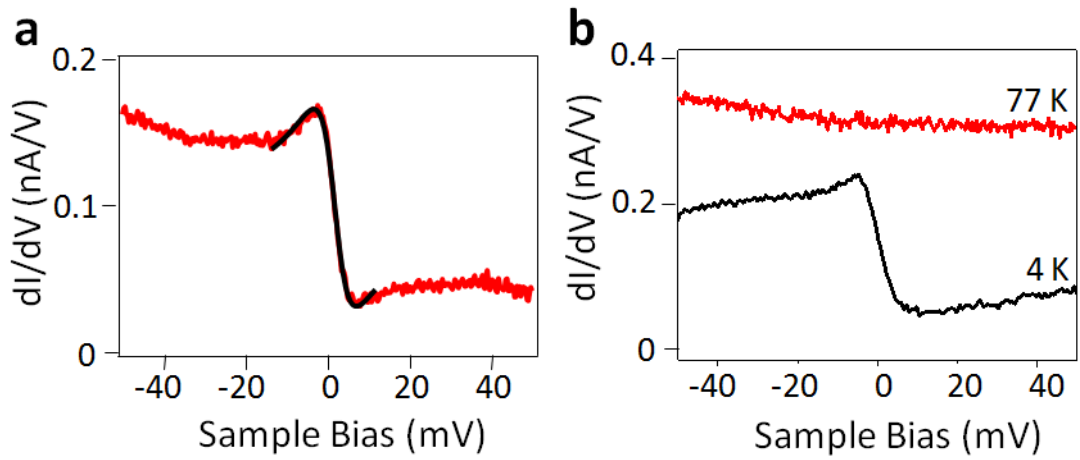

**Supplementary Figure 8 | a,** A typical fitting for determining $q$ and $T_K$. The red curve is the experimental dI/dV of the MnPc/Au(111) measured at the center of the molecule, showing a Kondo resonance near the Fermi level. The black curve is the Fano fitting. **b,** *dI/dV* spectra of the monolayer MnPc molecules at 4.2 K and 77 K, respectively, showing clearly quenching of Kondo effect by increasing temperature.

## 7. Reversible spin switching in molecular array

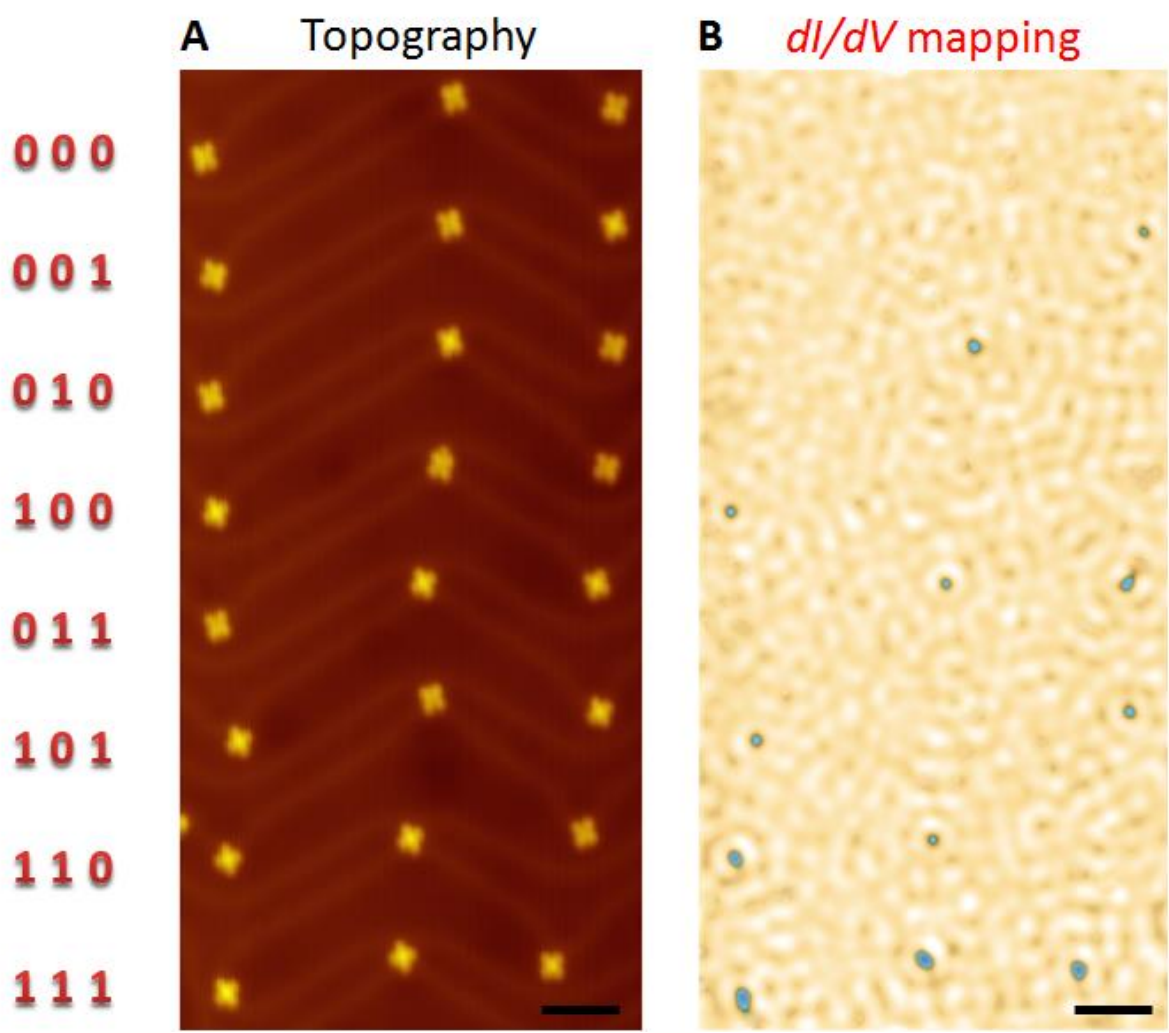

**Supplementary Figure 9 |** Topography and dI/dV mapping of a 3×8 molecular array with digital information corresponding to molecular Kondo ON/OFF states. **a,** Topography of a 3 × 8 molecular array, with encoded spin information that can be clearly revealed in the simultaneously acquired conductance mapping. Scanning bias: -0.2 V; Tunneling current: 10 pA; Scale bar, 5 nm. **b,** The dI/dV mapping at 4 mV acquired simultaneously with **a**. Kondo ON is defined as "1" state and Kondo OFF is defined as "0" state. According to this notation, a series of digitized states from "0 0 0" to "1 1 1" can be designed and created based on the spin state of individual molecules within the array.

# 8. The reason why we did not consider the vdW interaction in this calculations

In our formerly published paper[8], we discussed in detail the reason why we didn't include van der Waals interaction in our calculation. The vdW interaction induced a close molecule-substrate distance and then influenced the electronic structure at the interface. However, even though the dispersion corrected DFT method (DFT-D), in which vdW interaction is explicitly incorporated by using dispersion force field[9], performs fairly well in π-π packing systems[10], it overbinds the molecules to the metal substrate[11-14], and sometimes overestimates the binding energy with an error that is larger than the underestimates of a PBE functional[13]. One example in which the dispersion corrected DFT (DFT-D) method might give a wrong conclusion is the CoPc/Cu(111) system. In this system, a modified DFT-D method gave a stronger binding energy than did LDA[15]. The recently developed vdW-DF method[16] sometimes did not perform well in a molecule-metal interface either. According to published results, vdW-DF calculations predicted a binding distance between the aromatic molecules and Cu(111) substrate to be much larger than the experimental results[17,18]. At the same time, the traditional GGA functional works well in some MPc-metal systems. Baran *et al.*'s calculated results of CoPc(SnPc)/Ag(111) using the PBE-GGA functional showed excellent agreement with experiments. Taking the SnPc/Ag(111) system as an example, the calculated 3.7 Å Pc-surface distance was relatively close to the experimental result (3.6 Å)[19]. The Sn-surface distance also fit very well. In the PW91 functional, there is overbinding coming from the exchange part. Zhang et al. shows this will make the PW91functional occasionally work fairly well in a vdW-dominated system[20]. Therefore, we used a PW91-GGA functional to make the systematic study.

## Supplementary References: